\documentclass[journal]{IEEEtran}
\usepackage{calc}

\usepackage{multirow}
\usepackage{cite}
\usepackage{graphicx,subfigure}
\usepackage{psfrag}
\usepackage{amsmath,amssymb}
\usepackage{color}
\usepackage{tikz}
\usepackage{pgfplots}
\usepackage{mathtools}
\DeclarePairedDelimiter{\ceil}{\lceil}{\rceil}

\usetikzlibrary{snakes}

\newenvironment{proof}{\begin{IEEEproof}}{\end{IEEEproof}}

\DeclareMathOperator*{\dotleq}{\overset{.}{\leq}}
\DeclareMathOperator*{\dotgeq}{\overset{.}{\geq}}

\newtheorem{theorem}{Theorem}
\newtheorem{corollary}{Corollary}[theorem]

\newtheorem{lemma}{Lemma}

\newtheorem{example}{Example} 

\newcommand{\bit}{\begin{itemize}}
\newcommand{\eit}{\end{itemize}}

\newcommand{\bc}{\begin{center}}
\newcommand{\ec}{\end{center}}

\newcommand{\ba}{\begin{array}}
\newcommand{\ea}{\end{array}}

\newcommand{\beq}{\begin{equation}}
\newcommand{\eeq}{\end{equation}}

\newcommand{\beqn}{\begin{equation*}}
\newcommand{\eeqn}{\end{equation*}}

\newcommand{\bean}{\begin{eqnarray*}}
\newcommand{\eean}{\end{eqnarray*}}
\newcommand{\bea}{\begin{eqnarray}}
\newcommand{\eea}{\end{eqnarray}}

\def\C{\mathbb{C}}
\def\E{\mathbb{E}}

\def\gv{\boldsymbol{g}}
\def\hv{\boldsymbol{h}}

\def\xv{\boldsymbol{x}}

\newtheorem{remark}{Remark}

\begin{document}
\sloppy

\title{Fundamental Limits of Cache-Aided Wireless BC: Interplay of Coded-Caching and CSIT Feedback}
\author{Jingjing Zhang and Petros Elia
\thanks{The authors are with the Mobile Communications Department at EURECOM, Sophia Antipolis, 06410, France (email: jingjing.zhang@eurecom.fr, elia@eurecom.fr).
The work of Petros Elia was supported by the European Community's Seventh Framework Programme (FP7/2007-2013) / grant agreement no.318306 (NEWCOM\#), and from the ANR Jeunes Chercheurs project ECOLOGICAL-BITS-AND-FLOPS.}
\thanks{An initial version of this paper has been reported as Research Report No. RR-15-307 at EURECOM, August 25, 2015, http://www.eurecom.fr/publication/4723 as well as was uploaded on arxiv in November 2015.}
}


\maketitle

\thispagestyle{empty}

\begin{abstract}
Building on the recent coded-caching breakthrough by Maddah-Ali and Niesen, the work here considers the $K$-user cache-aided wireless multi-antenna (MISO) symmetric broadcast channel (BC) with random fading and imperfect feedback, and analyzes the throughput performance as a function of feedback statistics and cache size. In this setting, our work identifies the optimal cache-aided degrees-of-freedom (DoF) within a factor of 4, by identifying near-optimal schemes that exploit the new synergy between coded caching and delayed CSIT, as well as by exploiting the unexplored interplay between caching and feedback-quality.

The derived limits interestingly reveal that --- the combination of imperfect quality current CSIT, delayed CSIT, and coded caching, guarantees that --- the DoF gains have an initial offset defined by the quality of current CSIT, and then that the additional gains attributed to coded caching are exponential, in the sense that any linear decrease in the required DoF performance, allows for an exponential reduction in the required cache size. 
\end{abstract}

\section{Introduction\label{sec:intro}}
Recent work by \cite{MN14} explored --- for the single-stream broadcast setting --- how careful caching of content at the receivers, and proper encoding across different users' requested data, can allow for higher communication rates. The key idea was to use coding in order to create multicast opportunities, even if the different users requested different data content.
This \emph{coded caching} approach --- which went beyond storing popular content closer to the user --- involved two phases; the placement phase (during off peak hours) and the delivery phase (during peak hours). During the placement phase, content that was predicted to be popular (a library of commonly requested files), was coded and placed across user's caches.  During the delivery phase --- which started when users requested specific files from the predicted library of files --- the transmitter encoded across different users' requested data content, taking into consideration the requests and the existing cache contents. This approach --- which translated to efficient interference removal gains that were termed as `coded-caching gains' --- was shown in \cite{MN14} to provide substantial performance improvement that far exceeded the `local' caching gains from the aforementioned traditional `data push' methods that only pre-store content at local caches.

Our interest here is to explore coded caching, not in the original single-stream setting in~\cite{MN14}, but rather in the feedback-aided multi-antenna wireless BC.
This wireless and multi-antenna element now automatically brings to the fore a largely unexplored and involved relationship between coded caching and CSIT-type feedback quality. This relationship carries particular importance because both CSIT and coded caching are powerful and crucial ingredients in handling interference, because they are both hard to implement individually, and because their utility is affected by one another (often adversely, as we will see). Our work tries to understand how CSIT and caching resources jointly improve performance, as well as tries to shed some light on the interplay between coded caching and feedback.

\subsubsection{Motivation for the current work}
A main motivation in~\cite{MN14} and in subsequent works, was to employ coded caching to remove interference. Naturally, in wireless networks, the ability to remove interference is very much linked to the quality and timeliness of the available feedback, and thus
any attempt to further our understanding of the role of coded caching in these networks, stands to benefit from understanding the interplay between coded caching and (variable quality) feedback. This joint exposition becomes even more meaningful when we consider the connections that exist between feedback-usefulness and cached side-information at receivers, where principally the more side information receivers have, the less feedback information the transmitter might need.

This approach is also motivated by the fact that feedback is hard to get in a timely manner, and hence is typically far from ideal and perfect.
Thus, given the underlying links between the two, perhaps the strongest reason to jointly consider coded caching and feedback, comes from the prospect of using coded caching to alleviate the constant need to gather and distribute CSIT, which --- given typical coherence durations --- is an intensive task that may have to be repeated hundreds of times per second during the transmission of content. This suggests that content prediction of a predetermined library of files during the night (off peak hours), and a subsequent caching of parts of this library content again during the night, may go beyond boosting performance, and may in fact offer the additional benefit of alleviating the need for prediction, estimation, and communication of CSIT during the day, whenever requested files are from the library.
Our idea of exploring the interplay between feedback (timeliness and quality) and coded caching, hence draws directly from this attractive promise that content prediction, once a day, can offer repeated and prolonged savings in CSIT.

\subsection{Cache-aided broadcast channel model}
\subsubsection{$K$-user BC with pre-filled caching}
In the symmetric $K$-user multiple-input single-output (MISO) broadcast channel of interest here, the $K$-antenna transmitter, communicates to $K$ single-antenna receiving users. The transmitter has access to a library of $N\geq K$ distinct files $W_1,W_2, \dots, W_N$, each of size $|W_n| = f$ bits. Each user $k \in \{1,2,\dots,K\}$ has a cache $Z_k$, of size $|Z_k| = Mf$ bits, where naturally $M \leq N$. Communication consists of the aforementioned \emph{content placement phase} and the \emph{delivery phase}. During the placement phase --- which usually corresponds to communication during off-peak hours --- the caches $Z_1, Z_2, \dots, Z_K$ are pre-filled with content from the $N$ files $\{W_n\}_{n=1}^{N}$. The delivery phase commences when each user $k$ requests from the transmitter, any \emph{one} file $W_{R_k}\in \{W_n\}_{n=1}^{N}$, out of the $N$ library files. Each file can be requested with equal probability. Upon notification of the users' requests, the transmitter aims to deliver the (remaining of the) requested files, each to 
their intended receiver, and the challenge is to do so over a limited (delivery phase) duration $T$.

For each transmission, the received signals at each user $k$, will be modeled as
\begin{align}
y_{k}=\hv_{k}^{T} \xv + z_{k}, ~~ k = 1, \dots, K
\end{align}
where $\xv\in\mathbb{C}^{K\times 1}$ denotes the transmitted vector satisfying a power constraint $\E(||\xv||^2)\leq P$, where $\hv_{k}\in\mathbb{C}^{K\times 1}$ denotes the channel of user $k$ in the form of the random vector of fading coefficients that can change in time and space, and where $z_{k}$ represents unit-power AWGN noise at receiver $k$.
\begin{figure}[t!]
  \centering
\includegraphics[width=0.8\columnwidth]{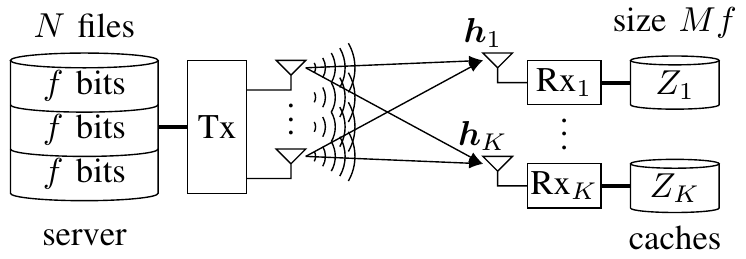}
\caption{Cache-aided $K$-user MISO BC.}
\label{fig:model}
\end{figure}
At the end of the delivery phase, each receiving user $k$ combines the received signal observations $y_{k}$ --- accumulated during the delivery phase --- with the fixed information in their respective cache $Z_k$, to reconstruct their desired file $W_{R_k}$.

\subsection{Coded caching and CSIT-type feedback}
Communication also takes place in the presence of channel state information at the transmitter. CSIT-type feedback is crucial in handling interference, and can thus substantially reduce the resulting duration $T$ of the delivery phase. This CSIT is typically of imperfect-quality as it is hard to obtain in a timely and reliable manner. In the high-SNR (high $P$) regime of interest, this current-CSIT quality is concisely represented in the form of the normalized quality exponent \cite{YKGY:12d}\cite{CE:13it}
\begin{align}
\alpha & := -\lim_{P \rightarrow \infty} \frac{\log \E[||{\hv_{k}}-{\hat \hv_{k}}||^2]}{\log P}, ~k\in \{1,\dots,K\}
\end{align}
where ${\hv_{k}}-{\hat \hv_{k}}$ denotes the estimation error between the current CSIT estimate ${\hat \hv_{k}}$ and the estimated channel ${\hv_{k}}$. The range of interest\footnote{In the high SNR regime of interest here, $\alpha=0$ corresponds to having essentially no current CSIT (cf.~\cite{DJ:14}), while having $\alpha = 1$ corresponds (again in the high SNR regime) to perfect and immediately available CSIT (cf.~\cite{Caire+:10m}).} is $\alpha\in[0,1]$. We also assume availability of delayed CSIT (as in for example \cite{MAT:11c}, as well as in a variety of subsequent works~\cite{YKGY:12d,CE:13it,GJ:12o,CE:12d,KYG:13,CYE:13isit,VV:09,TJSP:12,LH:12,HC:13}, see also \cite{VV:11t,AGK:11o,Lee2012,Tandon2012b} as well as \cite{TAV:2015,BW:2015,LTA:2015}) where now the delayed estimates of any channel, can be received without error but with arbitrary delay, even if this delay renders this CSIT completely obsolete. As it is argued in \cite{YKGY:12d}, this mixed CSI model (partial current CSIT, and delayed CSIT) nicely captures different realistic settings that might involve channel correlations and an ability to improve CSI as time progresses. This same CSI model is particularly well suited for our caching-related setting here, because it explicitly reflects two key ingredients that are directly intertwined with coded caching; namely, feedback timeliness and feedback quality.


In terms of caching, we will consider the normalized
\begin{align} \label{eq:gamma1}
\gamma := \frac{M}{N}
\end{align}
as well as the cumulative
\begin{align}
\Gamma := \frac{KM}{N} = K\gamma.
\end{align}
The latter simply means that the sum of the sizes of the caches across all users, is $\Gamma$ times the volume of the $N$-file library. As in~\cite{MN14}, we will consider the case where $\Gamma = \{1,2,\cdots K\}$.

\paragraph{Intuitive links between $\alpha$ and $\gamma$} As we will see, $\alpha$ is not only linked to the performance --- where a higher $\alpha$ allows for better interference management and higher performance over the wireless delivery link --- but is also linked to caching; after all, the bigger the $\gamma$, the more side information the receivers have, the less interference one needs to handle (at least in symmetric systems), and the smaller the $\alpha$ that is potentially needed to steer interference. This means that principally, a higher $\gamma$ implies that more common information needs to be transmitted, which may (in some cases) diminish the utility of feedback which primarily aims to facilitate the opposite which is the transmission of private information. It is for example easy to see (we will see this later) that in the presence of $\Gamma = K-1$, there is no need for CSIT in order to achieve the optimal performance.


\subsection{Measures of performance in current work}
As in \cite{MN14}, the measure of performance here is the duration $T$ --- in time slots, per file served per user --- needed to complete the delivery process, \emph{for any request}. The wireless link capabilities, and the time scale, are normalized such that one time slot corresponds to the optimal amount of time it would take to communicate a single file to a single receiver, had there been no caching and no interference.
As a result, in the high $P$ setting of interest --- where the capacity of a single-user MISO channel scales as $\log_2(P)$ --- we proceed to set
\begin{align}\label{eq:f}
f = \log_2(P)
\end{align} which guarantees that the two measures of performance, here and in \cite{MN14}, are the same and can thus be directly compared\footnote{
We note that setting $f = \log_2(P)$ is simply a normalization of choice, and does not carry a `forced' relationship between SNR and file sizes. The essence of the derived results would remain the same for any other non-trivial normalization.}.

A simple inversion leads to the equivalent measure of the per-user DoF
\begin{align}  \label{eq:TtoDoF}
d(\gamma,\alpha)=\frac{1-\gamma}{T}
\end{align}
which captures the joint effect of coded caching and feedback\footnote{The DoF measure is designed to exclude the benefits of having some content already available at the receivers (local caching gain), and thus to limit the DoF between 0, and the interference free optimal DoF of 1.}.


\subsection{Notation and assumptions}

\subsubsection{Notation}
We will use the notation $H_n := \sum_{i=1}^{n} \frac{1}{i}$,
to represent the $n$-th harmonic number, and we will use $\epsilon_n := H_n-\log (n)$ to represent its logarithmic approximation error, for some integer $n$. We remind the reader that $\epsilon_n$ decreases with $n$, and that $\epsilon_\infty :=\lim \limits_{n \rightarrow \infty} H_n -  \log (n) $ is approximately $0.5772$.
$\mathbb{Z}$ will represent the integers, $\mathbb{Z}^{+}$ the positive integers, $\mathbb{R}$ the real numbers, $\binom{n}{k}$ the $n$-choose-$k$ operator, and $\oplus$ the bitwise XOR operation. We will use $[K]:= \{1,2,\cdots,K\}$. If $\psi$ is a set, then $|\psi|$ will denote its cardinality. For sets $A$ and $B$, then $A \backslash B$ denotes the difference set.
Complex vectors will be denoted by lower-case bold font. We will use $||\xv||^2$ to denote the magnitude of a vector $\xv$ of complex numbers. For a transmitted vector $\xv$, we will use $\text{dur}(\xv)$ to denote the transmission duration of that vector. For example, having $\text{dur}(\xv) = \frac{1}{10}T$ would simply mean that the transmission of vector $\xv$ lasts one tenth of the delivery phase.
In our high-$P$ setting of interest, we will also use $\doteq$ to denote \emph{exponential equality}, i.e., we will write $g(P)\doteq P^{B}$ to denote $\displaystyle\lim_{P\to\infty}\frac{\log g(P)}{\log P}=B$.  Similarly $\dotgeq$ and $\dotleq$ will denote exponential inequalities.  Logarithms are of base~$e$, unless we use $\log_2(\cdot)$ which will represent a logarithm of base~2.

\subsubsection{Main assumptions}
Throughout this work, we assume availability of current CSIT with some quality $\alpha$, of delayed CSIT (D-CSIT), as well as ask that each receiver knows their own channel perfectly. We also adhere to the common convention (see for example~\cite{MAT:11c}) of assuming perfect and global knowledge of delayed channel state information at the receivers (delayed global CSIR), where each receiver must know (with delay) the CSIR of (some of the) other receivers. We will assume that the entries of \emph{each specific} estimation error vector are i.i.d. Gaussian.
For the outer (lower) bound to hold, we will make the common assumption that the current channel state must be independent of the previous channel-estimates and estimation errors, \emph{conditioned on the current estimate} (there is no need for the channel to be i.i.d. in time). We will make the assumption that the channel is drawn from a continuous ergodic distribution
such that all the channel matrices and all their sub-matrices are full rank almost surely.
We also make the soft assumption that the transmitter \emph{during the delivery phase} is aware of the feedback statistics. We note though that, while our main scheme assumes knowledge of $\alpha$ during the caching phase, most results will be the outcome of a simpler scheme that does not require knowledge of $\alpha$ during this caching phase. Removing this assumption entails, for $\alpha>0$, a performance penalty which is small. 

\subsection{Prior work}%
%

\nocite{KPR:99,BR:05,BK:06,BGW:10}

The benefits of coded caching on reducing interference and improving performance, were revealed in the seminal work by Maddah-Ali and Niesen in~\cite{MN14} who considered a caching system where a server is connected to multiple users through a shared link, and designed a novel caching and delivery method that jointly offers a multicast gain that helps mitigate the link load, and which was proven to have a gap from optimal that is at most 12. This work was subsequently generalized in different settings, which included the setting of different cache sizes for which Wang et al. in \cite{WLTL:15} developed a variant of the algorithm in ~\cite{MN14} which achieves a gap of at most 12 from the information theoretic optimal. Other extensions included the work in~\cite{MND13} by Maddah-Ali and Niesen who considered the setting of decentralized caching where the achieved performance was shown to be comparable to that of the centralized case~\cite{MN14}, despite the lack of coordination in content placement. For the same original single-stream setting of~\cite{MN14}, the work of Ji et al. in~\cite{JTLC:14} considered a scenario where users make multiple requests each, and proposed a scheme that has a gap to optimal that is less than 18. Again for the setting in~\cite{MN14}, the work of Ghasemi and Ramamoorthy in~\cite{HA:2015}, derived tighter outer (lower) bounds that improve upon existing bounds, and did so by recasting the bound problem as one of optimally labeling the leaves of a directed tree. Further work can be found in~\cite{WLG:15} where Wang et al. explored the interesting link between caching and distributed source coding with side information. Interesting conclusions are also drawn in the work of Ajaykrishnan et al. in~\cite{APPV:15}, which revealed that the effectiveness of caching in the single stream case, is diminished when $N$ approaches and exceeds $K^2$.

Deviating from single-stream error free links, different works have considered the use of coded caching in different wireless networks, without though particular consideration for CSIT feedback quality. For example, work by Huang et al. in~\cite{HuangWDY015}, considered a cache-aided wireless fading BC where each user experiences a different link quality, and proposed a suboptimal communication scheme that is based on time- and frequency-division and power- and bandwidth-allocation, and which was evaluated using numerical simulations to eventually show that the produced throughput decreases as the number of users increases. Further work by Timo and Wigger in~\cite{TW:15} considered an erasure broadcast channel and explored how the cache-aided system efficiency can improve by employing unequal cache sizes that are functions of the different channel qualities. Another work can be found in~\cite{MN:15isit} where Maddah-Ali and Niesen studied the wireless interference channel where each transmitter has a local cache, and showed distinct benefits of coded caching that stem from the fact that content-overlap at the transmitters allows effective interference cancellation.

Different work has also considered the effects of caching in different non-classical channel paradigms. One of the earlier such works that focused on practical wireless network settings, includes the work by Golrezaei et al. in~\cite{GSDMC:12}, which considered a downlink cellular setting where the base station is assisted by helper nodes that jointly form a wireless distributed caching network (no coded caching) where popular files are cached, resulting in a substantial increase to the allowable number of users by as much as $400 - 500\%$. In a somewhat related setting, the work in~\cite{PBKD:15} by Perabathini et al. accentuated the energy efficiency gains from caching.
Further work by Ji et al. in~\cite{JWTLCEL:15} derived the limits of so-called combination caching networks in which a source is connected to multiple user nodes through a layer of relay nodes, such that each user node with caching is connected to a distinct subset of the relay nodes. Additional work can also be found in~\cite{NSW:12} where Niesen et al. considered a cache-aided network where each node is randomly located inside a square, and it requests a message that is available in different caches distributed around the square. Further related work on caching can be found in \cite{BBD:15,MCOFBJ:14,HKD:14,HKS:15,SJTLD:15,JTLC:14}.

Work that combines caching and feedback considerations in wireless networks, has only just recently started.
A reference that combines these, can be found in~\cite{DBAD:15} where Deghel et al. considered a MIMO interference channel (IC) with caches at the transmitters. In this setting, whenever the requested data resides within the pre-filled caches, the data-transfer load of the backhaul link is alleviated, thus allowing for these links to be instead used for exchanging CSIT that supports interference alignment. An even more recent concurrent work can be found in~\cite{GKY:15} where Ghorbel et al. studied the capacity of the cache-enabled broadcast packet erasure channel with ACK/NACK feedback. In this setting, Ghorbel et al. cleverly showed --- interestingly also using a retrospective type algorithm, this time by Gatzianas et al. in~\cite{GGT:13} --- how feedback can improve performance by informing the transmitter when to resend the packets that are not received by the intended user and which are received by unintended users, thus allowing for multicast opportunities. The first work that considers the actual interplay between coded caching and CSIT quality, can be found in~\cite{ZFE:15} which considered the easier problem of how the optimal cache-aided performance (with coded caching), can be achieved with reduced quality CSIT.

\subsection{Outline and contributions}

In Section~\ref{sec:mainResults}, Lemma~\ref{lem:outer}, we offer a lower bound for the optimal $T^*(\gamma,\alpha)$.
Then in Theorem~\ref{thm:bigGamma} we calculate the achievable $T(\gamma,\alpha)$, for $\Gamma \in \{1, 2, \cdots , K\}$, $\alpha \in[0,1]$, and prove it to be less than four times the optimal, thus identifying the optimal $T^*(\gamma,\alpha)$ within a factor of 4. A simpler expression for $T$ (again within a factor of $4$ from optimal), and its corresponding per-user DoF, are derived in Theorem~\ref{thm:bigGamma}, while a simple approximation of these is derived in Corollary~\ref{cor:LargeGammaLogApprox}, where we see that the per-user DoF takes the form $d(\gamma,\alpha) = \alpha + (1-\alpha) \frac{1-\gamma}{\log{\frac{1}{\gamma}}}$, revealing that even a very small $\gamma = e^{-G}$ can offer a substantial DoF boost $
d(\gamma = e^{-G},\alpha) - d(\gamma = 0,\alpha) \approx (1-\alpha)\frac{1}{G}.$

In Section~\ref{sec:CacheAidedCSIT} we discuss practical implications. In~Corollary~\ref{cor:alphaGainTot1} we describe the savings in current CSIT that we can have due to coded caching, while in Corollary~\ref{cor:alphaThreshold2GtotLarge} we quantify the intuition that, in the presence of coded-caching, there is no reason to improve CSIT beyond a certain threshold quality.
Furthermore in Section~\ref{sec:vanishingFractionCSIT} we show how cache-aided communications can utilize a vanishingly-small portion of D-CSIT compared to traditional D-CSIT schemes, simply because caching helps `skip' the parts of the schemes that require the highest D-CSIT load.

In Section~\ref{sec:schemeAlphaBigGamma} we present the caching-and-delivery schemes, which build on the interesting connections between MAT-type retrospective transmission schemes (cf.~\cite{MAT:11c}) and coded caching. The caching part is modified from~\cite{MN14} to essentially \emph{`fold'} (linearly combine) the different users' data into multi-layered blocks, in a way such that the subsequent transmission algorithm (which employs parts of the QMAT algorithm in \cite{KGZE:16}) is suited to efficiently unfold these. The caching and transmission algorithms are calibrated so that the caching algorithm --- which is modified from that in~\cite{MN14} to adapt the caching redundancy to $\alpha$ --- creates the same multi-destination delivery problem that is efficiently solved by the last stages of the QMAT scheme.  
Section~\ref{sec:additionalProofs} in the Appendix presents the outer bound proof, and the proof for the gap to optimal.

\section{Throughput of cache-aided BC as a function of CSIT quality and caching resources\label{sec:mainResults}}
The following results hold for the $(K,M,N,\alpha)$ cache-aided $K$-user wireless MISO BC with random fading, $\alpha\in [0,1]$ and $N\geq K$, where $\gamma = \frac{M}{N}$ and $\Gamma = K\gamma$. We begin with an outer bound (lower bound) on the optimal $T^*$.

\vspace{3pt}
\begin{lemma}\label{lem:outer}
The optimal $T^*$ for the $(K,M,N,\alpha)$ cache-aided $K$-user MISO BC, is lower bounded as
\begin{align}
T^*(\gamma,\alpha) \geq \mathop {\text{max}}\limits_{s\in \{1, \dots, \lfloor \frac{N}{M} \rfloor \}} \frac{1}{(H_s \alpha+1-\alpha)} (H_s -\frac{Ms}{\lfloor \frac{N}{s} \rfloor}).
\end{align}
\end{lemma}
\vspace{3pt}
\begin{proof}
The proof is presented in Section~\ref{sec:lower} and it uses the bound from Lemma~\ref{lem:lowerSecond} whose proof can be found in Section~\ref{sec:lowerSecond}.
\end{proof}
\vspace{3pt}

\subsection{Achievable throughput of the cache-aided BC}
The following identifies, up to a factor of 4, the optimal $T^*$, for all $\Gamma \in \{1, 2, \cdots , K\}$ (i.e., $M\in \frac{N}{K}\{1,\cdots,K\}$). The result uses the expression \begin{align} \label{eq:alphaBreak}
\alpha_{b,\eta} = \frac{\eta-\Gamma}{\Gamma(H_K-H_\eta-1)+\eta},  \ \eta = \ceil{\Gamma},\dots,K-1. \end{align} Note that the above does not hold for $\Gamma = K$, as this would imply no need for delivery.

\vspace{3pt}
\begin{theorem} \label{thm:bigGammaBest}
In the $(K,M,N,\alpha)$ cache-aided MISO BC with $N$ files, $K\leq N$ users, $\Gamma \in \{1, 2, \cdots , K\}$, and for $\eta = \arg\max_{\eta{'}\in [\Gamma,K-1]\cap \mathbb{Z}} \{\eta{'} \ : \ \alpha_{b,\eta'}\leq \alpha\}$, then
\begin{align}\label{eq:gammabigBest}
T = \max\{1-\gamma, \frac{(K-\Gamma)(H_K-H_\eta)}{(K-\eta)+\alpha(\eta+K(H_K-H_\eta-1))}\}
\end{align} is achievable and always has a gap-to-optimal that is less than 4, for all $\alpha,K$. For $\alpha \geq \frac{K(1-\gamma)-1}{(K-1)(1-\gamma)} $, $T$ is optimal.
\end{theorem}
\vspace{3pt}
\begin{proof}
The caching and delivery scheme that achieves the above performance is presented in Section~\ref{sec:schemeAlphaBigGamma}, while the corresponding gap to optimal is bounded in Section~\ref{sec:gapCalculation}.
\end{proof}
\vspace{3pt}

The above is achieved with a general scheme whose caching phase is a function of $\alpha$. We will henceforth consider a special case ($\eta =\Gamma$) of this scheme, which provides similar performance (it again has a gap to optimal that is bounded by 4), simpler expressions, and has the practical advantage that the caching phase need not depend on the CSIT statistics $\alpha$ of the delivery phase. For this case, we can achieve the following performance.

\vspace{3pt}
\begin{theorem} \label{thm:bigGamma}
In the $(K,M,N,\alpha)$ cache-aided MISO BC with $\Gamma \in \{1, 2, \cdots , K\}$,
\begin{align}\label{eq:gammabig}
T = \frac{(1-\gamma)(H_K-H_{\Gamma})}{\alpha(H_K-H_{\Gamma})+(1-\alpha)(1-\gamma)}
\end{align}
is achievable and has a gap from optimal
\begin{align}\label{eq:gap2}
\frac{T}{T^*}<4
\end{align}
that is less than 4, for all $\alpha,K$. Thus the corresponding per-user DoF takes the form
\begin{align}\label{eq:gammabigDoF}
d(\gamma,\alpha) = \alpha + (1-\alpha)\frac{1-\gamma}{H_K-H_\Gamma}.
\end{align}
\end{theorem}
\vspace{3pt}
\begin{proof}
The scheme that achieves the above performance will be described later on as a special (simpler) case of the scheme corresponding to Theorem~\ref{thm:bigGammaBest}. The corresponding gap to optimal is bounded in Section~\ref{sec:gapCalculation}.
\end{proof}
\vspace{3pt}

%
The following corollary describes the above achievable $T$, under the logarithmic approximation $H_n\approx\log (n)$. The presented expression is exact in the large $K$ setting\footnote{For large $K$, this approximation
$\frac{H_K-H_{\Gamma}}{\log(\frac{1}{\gamma})} = 1$ is tight for any \emph{fixed} $\gamma$.} where
$\frac{H_K-H_{\Gamma}}{\log(\frac{1}{\gamma})} = 1$.
\vspace{3pt}
\begin{corollary} \label{cor:LargeGammaLogApprox}
Under the logarithmic approximation $H_n\approx\log (n)$, the derived $T$ takes the form
\begin{align}\label{eq:gammabigApprox}
T(\gamma,\alpha) = \frac{(1-\gamma)\log(\frac{1}{\gamma})}{\alpha\log(\frac{1}{\gamma})+(1-\alpha)(1-\gamma)}
\end{align}
and the derived DoF takes the form
\begin{align}\label{eq:gammabigApproxDoFLog}
d(\gamma,\alpha) = \alpha + (1-\alpha) \frac{1-\gamma}{\log{\frac{1}{\gamma}}}.
\end{align}
\end{corollary}
\vspace{3pt}
For the large $K$ setting, what the above suggests is that current CSIT offers an initial DoF boost of $d^*(\gamma=0,\alpha) = \alpha$ (cf.~\cite{KGZE:16}), which is then supplemented by a DoF gain
\[d(\gamma,\alpha) - d^*(\gamma=0,\alpha) \rightarrow (1-\alpha)\frac{1-\gamma}{\log(\frac{1}{\gamma})}\]
attributed to the synergy between delayed CSIT and caching~\footnote{We note that these interference-removal gains, particularly in the large $K$ regime, are not a result of extra performance boost directly from D-CSIT, because in the large $K$ setting, this latter performance boost is negligible (vanishes to zero) without caching.}. These synergistic gains (see also~\cite{ZEsynergy:16}) are accentuated for smaller values of $\gamma$, where we see an exponential effect of coded caching, in the sense that now a microscopic $\gamma = e^{-G}$ can offer a substantial DoF boost
\begin{align}
d(\gamma = e^{-G},\alpha) - d(\gamma = 0,\alpha) \approx (1-\alpha)\frac{1}{G}.
\end{align}

\begin{example} \label{ex:GapToOptimal}
In a MISO BC system with $\alpha = 0$, $K$ antennas and $K$ users, in the absence of caching, the optimal per-user DoF is $d^*(\gamma=0,\alpha=0) = 1/H_K$ (cf.~\cite{MAT:11c}) which vanishes to zero as $K$ increases. A DoF of $1/4$ can be guaranteed with $\gamma \approx \frac{1}{50}$ for all $K$, a DoF of $1/7$ with $\gamma \approx \frac{1}{1000}$, and a DoF of $1/11.7$ can be achieved with $\gamma \approx 10^{-5}$, again for all $K$.
\end{example}

\paragraph{Interplay between CSIT quality and coded caching in the symmetric MISO BC}
The derived form in~\eqref{eq:gammabigDoF} (and its approximation in \eqref{eq:gammabigApproxDoFLog}) nicely capture the synergistic as well as competing nature of feedback and coded caching. It is easy to see for example that the effect from coded-caching, reduces with $\alpha$ and is proportional to $1-\alpha$. This reflects the fact that in the symmetric MISO BC, feedback supports broadcasting by separating data streams, thus diminishing multi-casting by reducing the number of common streams.
In the extreme case when $\alpha = 1$, we see --- again for the symmetric MISO BC --- that the caching gains are limited to local caching gains\footnote{This conclusion is general (and not dependent on the specific schemes), because the used schemes are optimal for $\alpha = 1$. The statement holds because we can simply uniformly cache a fraction $\gamma$ of each file in each cache, and upon request, use perfect-CSIT to zero-force the remaining requested information, to achieve the optimal $T^*(\gamma,\alpha = 1) = 1-\gamma$, which leaves us with local (data push) caching gains only.}.

\section{Cache-aided CSIT reductions\label{sec:CacheAidedCSIT}}
We proceed to explore how coded caching can alleviate the need for CSIT.

\subsection{Cache-aided CSIT gains} 
%
To capture the cache-aided reductions on the CSIT load, let us consider
\begin{multline}
\label{eq:alphaGainCode2}
\bar{\alpha}(\gamma,\alpha) := \arg\min_{\alpha'}\{\alpha': (1-\gamma) T^*(\gamma=0,\alpha') \leq T(\gamma,\alpha)\}
\end{multline}
which is derived below in the form \[\bar{\alpha}(\gamma,\alpha) = \alpha + \delta_\alpha(\gamma,\alpha)\] for some $\delta_\alpha(\gamma,\alpha)$ that can be seen as the \emph{CSIT reduction due to caching} (from $\bar{\alpha}(\gamma,\alpha)$ to the operational $\alpha$). 

\vspace{3pt}

\begin{corollary}
\label{cor:alphaGainTot1}
In the $(K,M,N,\alpha)$ cache-aided MISO BC, then
\begin{align} \label{eq:alphaGainTotalGeneral}
\bar{\alpha}(\gamma,\alpha) = \alpha + \frac{(1-\alpha)(H_{K\gamma}-\gamma H_K)}{(H_K-1)(H_K-H_{K\gamma})}
\end{align}
is achievable, and implies a cache-aided CSIT reduction
\[\delta_\alpha(\gamma,\alpha) = \frac{(1-\alpha)(H_{K\gamma}-\gamma H_K)}{(H_K-1)(H_K-H_{K\gamma})}.\]
\end{corollary}

\vspace{3pt}
\begin{proof}
The proof is direct from Theorem~\ref{thm:bigGamma}.
\end{proof}

The above is made more insightful in the large $K$ regime, for which we have the following.

\vspace{3pt}
\begin{corollary} \label{cor:alphaGainTot1Asymptotic1}
In the $(K,M,N,\alpha)$ cache-aided MISO BC, then
\begin{align} \label{eq:alphaGainCodingAsymptotic1b}
\bar{\alpha}(\gamma,\alpha) = \alpha+(1-\alpha) \frac{1-\gamma}{\log(\frac{1}{\gamma})}
\end{align}
which implies CSIT reductions of
\[\delta_\alpha(\gamma,\alpha) = (1-\alpha)d(\gamma,\alpha = 0) = (1-\alpha) \frac{1-\gamma}{\log(\frac{1}{\gamma})}.\]
\end{corollary}
\vspace{3pt}
\begin{proof}
The proof is direct from the definition of $\bar{\alpha}(\gamma,\alpha)$ and from Theorem~\ref{thm:bigGamma}.
\end{proof}
\vspace{3pt}

Furthermore we have the following which quantifies the intuition that, in the presence of coded-caching, there is no reason to improve CSIT beyond a certain threshold quality. The following uses the definition in~\eqref{eq:alphaBreak}, and it holds for all $K$.

\vspace{3pt}

\begin{corollary} \label{cor:alphaThreshold2GtotLarge}
For any $\Gamma\in \{1,\dots,K\}$, then
\begin{align} \label{eq:alphaThreshold1}
T^*(\gamma,\alpha) = T^*(\gamma,\alpha = 1) = 1-\gamma
\end{align} holds for any
\begin{align}
\alpha \geq \alpha_{b,K-1} = \frac{K(1-\gamma)-1}{(K-1)(1-\gamma)}
\end{align}
which reveals that CSIT quality $\alpha = \alpha_{b,K-1}$ is the maximum needed, as it already offers the same optimal performance $T^*(\gamma,\alpha = 1)$ that would be achieved if CSIT was perfect.
\end{corollary}
\vspace{3pt}
\begin{proof}
This is seen directly from Theorem~\ref{thm:bigGammaBest} after noting that the achievable $T$ matches $T^*(\gamma,\alpha = 1) = 1-\gamma$.
\end{proof}
\vspace{3pt}

\paragraph{How much caching is needed to partially substitute current CSIT with delayed CSIT (using coded caching to `buffer' CSI)}
As we have seen, in addition to offering substantial DoF gains, the synergy between feedback and caching can also be applied to reduce the burden of acquiring current CSIT.
What the above results suggest is that a modest $\gamma$ can allow a BC system with D-CSIT to approach the performance attributed to current CSIT, thus allowing us to partially substitute current with delayed CSIT, which can be interpreted as an ability to buffer CSI. A simple calculation --- for the large-$K$ regime --- can tell us that
\[\gamma^{'}_{\alpha}:= \arg\min_{\gamma^{'}}\{\gamma^{'}:  d(\gamma^{'},\alpha = 0) \geq d^*(\gamma = 0,\alpha)\} = e^{-1/\alpha}\]
which means that $
\gamma^{'}_{\alpha}= e^{-1/\alpha}$ suffices to achieve --- in conjunction with delayed CSIT --- the optimal DoF performance $d^*(\gamma = 0,\alpha)$ associated to a system with delayed CSIT and $\alpha$-quality current CSIT.

\begin{example}
Let $K$ be very large, and consider a BC system with delayed CSIT and $\alpha$-quality current CSIT, where $\alpha = 1/5$. Then $\gamma^{'}_{\alpha = 1/5}= e^{-5} = 0.0067 \approx 1/150$ which means that
\[ d^*(\gamma = 0.0067,\alpha=0) \geq d^*(\gamma = 0,\alpha=1/5)
\]
which says that the same high-$K$ per-user DoF performance $d^*(\gamma = 0,\alpha=1/5)$, can be achieved by substituting all current CSIT with coded caching employing $\gamma\approx 1/150$.
\end{example}

\subsection{Vanishing fraction of delayed CSIT\label{sec:vanishingFractionCSIT}}
In the following we briefly explore how caching allows for a reduced D-CSIT load. We do so for the case of $\alpha = 0$. 

When $\alpha = 0$, the delivery scheme which we describe in Section~\ref{sec:schemeAlphaBigGamma}, draws directly from the MAT scheme~\cite{MAT:11c}. This scheme can have up to $K$ phases which are of decreasing time duration and which use a decreasing number of transmit antennas. Essentially each phase is lighter than the previous one, in terms of implementation difficulty. What we will see is that caching will allow us to bypass the first $\Gamma$ phases, which are the longest and most intensive, leaving us with the remaining $K-\Gamma$ communication phases that are easier to support with delayed feedback because they involve fewer transmissions, with fewer transmit antennas and to fewer users, and thus involve fewer D-CSIT scalars that must be communicated. 

In brief --- after normalization to account for the condition that each user receives a total of $\log_2(P)$ bits of data --- each phase $j = \Gamma+1,\Gamma+2,\dots,K$ will have a \emph{normalized} duration $T_j = \frac{1}{j}$. During each phase $j$, we will need to send D-CSIT that describes the channel vectors for $K-j$ users, and during this same phase the transmitted vectors will have support $K-j+1$ because only $K-j+1$ transmit antennas are active. Thus during phase $j$, there will be a need to send $T_j(K-j+1)(K-j)  = \frac{1}{j}(K-j+1)(K-j)$ D-CSIT scalars, and thus a need to send D-CSIT for up to a total of
\begin{align}
& L(\Gamma) = \sum_{j=\Gamma+1}^K \frac{1}{j}(K-j+1)(K-j) \nonumber \\ &=
(K^2+K)(H_K-H_{\Gamma}) -\frac{K(1-\gamma)(3K-K\gamma-1)}{2} \nonumber \end{align} channel scalars, while in the absence of caching (corresponding to $\Gamma = 0$), we will have to send D-CSIT on
\begin{align}
L(\Gamma = 0) & = \sum_{j=1}^K \frac{1}{j}(K-j+1)(K-j) \nonumber \\ & = (K^2+K)H_K -\frac{3K^2}{2} + \frac{K}{2}\nonumber \end{align} channel scalars.

\begin{figure}[t!]
  \centering
 \includegraphics[scale=0.6]{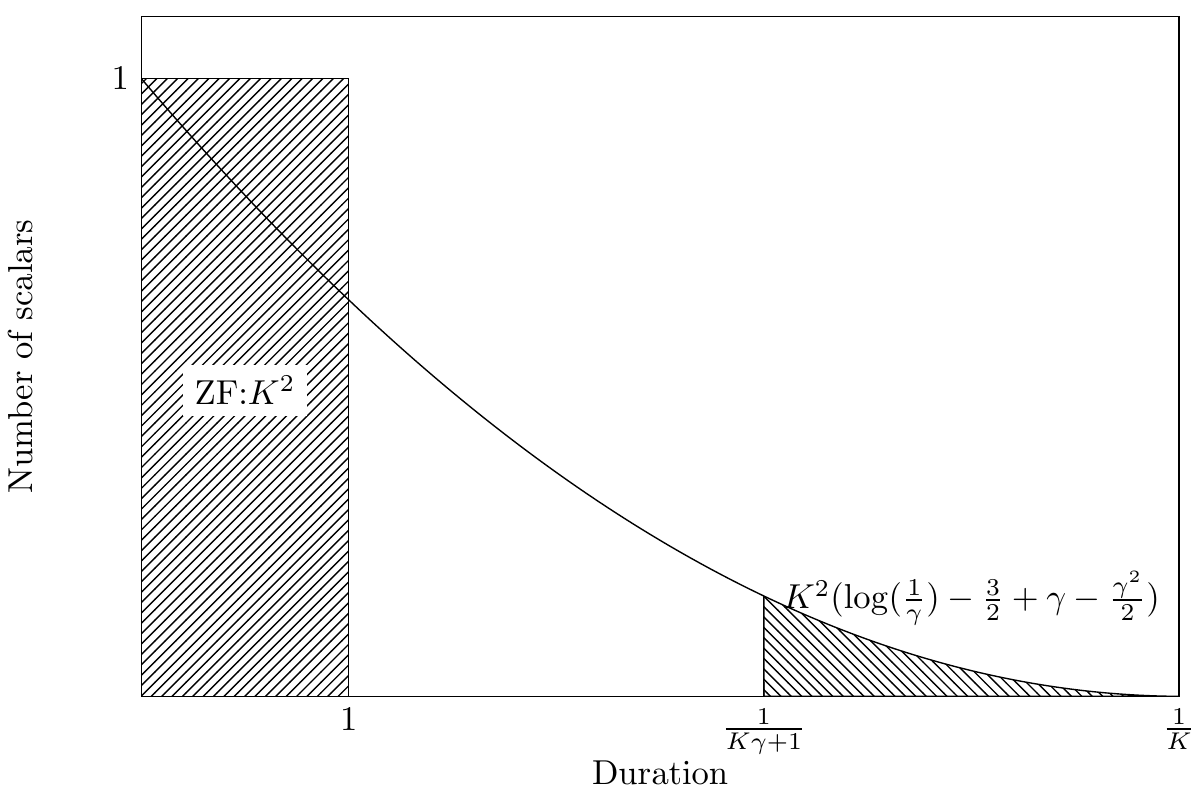}
\caption{Illustration of the vanishing fraction of D-CSIT cost, due to caching.}
\label{fig:vanishingDelayedCSIT2}\end{figure}

To reflect the frequency of having to gather D-CSIT, and to provide a fair comparison between different schemes of different performance that manage to convey different amounts of actual data to the users, we consider the measure $Q(\Gamma)$ that normalizes the above number $L(\Gamma)$ of full D-CSIT scalars, by the coherence period $T_c$ and by the total number of full data symbols sent. In our case, under the assumption that each user receives a total of $\log_2(P)$ bits, the total number of full data symbols sent is $K$, and thus we have
\begin{align}
Q(\Gamma) & = \frac{L(\Gamma)}{T_c K} \nonumber \\
 &=  \frac{(K^2+K)(H_K-H_{\Gamma}) -\frac{K(1-\gamma)(3K-K\gamma-1)}{2}}{T_c K}\nonumber
\end{align}
while without caching, we have
\begin{align}
Q(\Gamma = 0) & = \frac{L(\Gamma)}{T_c K}= \frac{(K+1)H_K - \frac{3}{2}K +\frac{1}{2}}{T_c}.\nonumber
\end{align}
Consequently we see that in the large $K$ limit,
\begin{align}
Q(\Gamma) \rightarrow \frac{K \bigl(\log(\frac{1}{\gamma})-\frac{3}{2}+2\gamma-2\gamma^2\bigr)}{T_c}\nonumber
\end{align}
\begin{align}
Q(\Gamma = 0) \rightarrow \frac{1}{T_c}K\log(K)\nonumber \end{align}
which implies that
\begin{align}
\lim_{K\rightarrow \infty} \frac{Q(\Gamma) }{Q(\Gamma = 0)} = 0 \nonumber
\end{align}
which in turn tells us that as $K$ increases, for any fixed $\gamma$, caching allows for a substantial reduction (down to a vanishingly small portion) from the original cost of D-CSIT. This is illustrated in Figure~\ref{fig:vanishingDelayedCSIT2}.

This reduction is important because retrospective delayed-feedback methods suffer from an increased cost of supporting their CSIT requirements (cf.~\cite{KC:12}) (albeit at the benefit of allowing substantial delays in the feedback mechanisms); after all, in the presence of perfect CSIT and zero forcing (no caching), the same cost is
\[Q_{ZF} = \frac{K^2}{T_c K} = \frac{K}{T_c}\] which gives that
\begin{align}
\lim_{K\rightarrow \infty} \frac{Q(\Gamma = 0) }{Q_{ZF}} = \infty\nonumber
\end{align}
which in turn verifies the above claim, and shows that the increase in the cost of supporting the D-CSIT (without caching) can be unbounded compared to ZF methods.
On the other hand, we see that
\begin{align}
\lim_{K\rightarrow \infty} \frac{Q(\Gamma) }{Q_{ZF}} = \log(\frac{1}{\gamma} - \frac{3}{2}+2\gamma-2\gamma^2)\nonumber
\end{align}
which means that \[\lim_{K\rightarrow \infty} \frac{Q(\Gamma) }{Q_{ZF}} <1, \ \gamma\geq \frac{1}{10}. \]
One interesting conclusion that comes out of this, is that caching can allow for full substitution of current CSIT (as we have seen above), with a very substantial reduction of the cost of D-CSIT as well, where for $\gamma\geq \frac{1}{10}$ this cost is even less than that of the very efficient ZF, which has to additionally deal though with harder-to-obtain current CSIT. This cost reduction is also translated into a reduction in the cost of disseminating global channel state information at the receivers (global CSIR), where each receiver must now know (again with delay that is allowed to be large) the CSIR of only a fraction of the other receivers.

\section{Cache-aided retrospective communications} \label{sec:schemeAlphaBigGamma}

We proceed to describe the communication scheme, and in particular the process of placement, folding-and-delivery, and decoding. In the end we calculate the achievable duration $T$.

The caching part is modified from~\cite{MN14} to \emph{`fold'} (linearly combine) the different users' data into multi-layered blocks, in a way such that the subsequent Q-MAT transmission algorithm (cf.~\cite{KGZE:16}) (specifically the last $K-\eta_\alpha$ ($\eta_\alpha\in\{\Gamma,\dots,K-1\}$) phases of the QMAT algorithm) can efficiently deliver these blocks. Equivalently the algorithms are calibrated so that the caching algorithm creates a multi-destination delivery problem that is the same as that which is efficiently solved by the last stages of the QMAT-type communication scheme. We henceforth remove the subscript in $\eta_\alpha$ and simply use $\eta$, where now the dependence on $\alpha$ is implied.

\subsection{Placement phase}
We proceed with the placement phase which modifies on the work of~\cite{MN14} such that when the CSIT quality $\alpha$ increases, the algorithm caches a decreasing portion from each file, but does so with increasing redundancy. The idea is that the higher the $\alpha$, the more private messages one can deliver directly without the need to multicast, thus allowing for some of the data to remain entirely uncached, which in turn allows for more copies of the same information across different users' caches.

Here each of the $N$ files $W_n, n = 1, 2, \ldots, N$ ($|W_n| = f$ bits) in the library, is split into two parts
\begin{align} \label{eq:splitCachedUncached}
W_n = (W_n^c, W_n^{\overline{c}})
\end{align}
where $W_n^c$ ($c$ for `cached') will be placed into one or more caches, while the content of $W_n^{\overline{c}}$ ($\overline{c}$ for `non-cached') will never be cached anywhere, but will instead be communicated --- using CSIT --- in a manner that causes manageable interference and hence does not necessarily benefit from coded caching.
The split is such that
\begin{align}\label{eq:WNcSize}
|W_n^c| = \frac{KMf}{N\eta}
\end{align}
where $\eta\in\{\Gamma,\dots,K-1\}$ is a positive integer, the value of which will be decided later on such that it properly regulates how much to cache from each $W_n$. Now for any specific $\eta$, we equally divide $W_n^c$ into
$\binom{K}{\eta}$ subfiles $\{W^c_{n,\tau}\}_{\tau \in \Psi_{\eta}} $,
\begin{align} \label{eq:WnTau}
W_n^c = \{W^c_{n,\tau}\}_{\tau \in \Psi_{\eta}}
\end{align}
where\footnote{We recall that in the above, $\tau$ and $W^c_{n,\tau}$ are sets, thus $|\tau|,|W^c_{n,\tau}|$ denote cardinalities; $|\tau| = \eta$ means that $\tau$ has $\eta$ different elements from $[K]$, while $|W^c_{n,\tau}|$ describes the size of $W^c_{n,\tau}$ in bits.} \begin{align}\label{eq:PsiEta} \Psi_{\eta}:= \{\tau \subset [K] \ : \  |\tau| = \eta\}\end{align} where each subfile has size
\begin{align} \label{eq:WnTauSize}
|W^c_{n,\tau}| = \frac{KMf}{N\eta\binom{K}{\eta}} = \frac{Mf}{N\binom{K-1}{\eta-1}} \ \text{bits}.\end{align}

Now drawing from \cite{MN14}, the caches are filled as follows
\begin{align}\label{eq:ZkFill1} Z_k=\{W^c_{n,\tau}\}_{n \in [N], \tau\in \Psi_{\eta}^{(k)}}\end{align}
where
\begin{align}\label{eq:PsiEta_k}
\Psi_{\eta}^{(k)} := \{\tau \in \Psi_{\eta} \ : \ k\in \tau\}.
\end{align}
Hence each subfile $W^c_{n, \tau}$ is stored in $Z_k$ as long as $k\in\tau$, which means that each $W^c_{n, \tau}$ (and thus each part of $W_n^c$) is repeated $\eta$ times in the caches. As $\eta$ increases with $\alpha$, this means that CSIT allows for a higher redundancy in the caches; instead of content appearing in $\Gamma$ different caches, it appears in $\eta\geq \Gamma$ caches instead, which will translate into multicast messages that are intended for more receivers.

\subsection{Data folding}
At this point, the transmitter becomes aware of the file requests $R_k, k=1,\dots,K$, and must now deliver each requested file $W_{R_k}$, by delivering the constituent subfiles $\{W^c_{R_k,\tau}\}_{\tau\in\Psi_{\eta} \backslash \Psi_{\eta}^{(k)}}$ as well as $W_{R_k}^{\overline{c}}$, all to the corresponding receiver $k$. We quickly recall that:
\begin{enumerate}
    \item  subfiles $\{W^c_{R_k,\tau}\}_{\tau \in \Psi_{\eta}^{(k)}} $ are already in $Z_k$;
    \item  subfiles $\{W^c_{R_k,\tau}\}_{\tau\in\Psi_{\eta} \backslash \Psi_{\eta}^{(k)}}$ are directly requested by user $k$, but are not cached in $Z_k$;
    \item  subfiles $ Z_k \backslash \{W^c_{R_k,\tau}\}_{\tau \in \Psi_{\eta}^{(k)}} =  Z_k \backslash W^c_{R_k}$ are cached in $Z_k$, are not directly requested by user $k$, but will be useful in removing interference.
\end{enumerate}

We assume the communication here has duration $T$. Thus for each $k$ and a chosen $\eta$, we split each subfile $W^c_{R_k,\tau}, \ \tau\in\Psi_{\eta} \backslash \Psi_{\eta}^{(k)}$ (each of size $|W^c_{R_k,\tau}| = \frac{Mf}{N\binom{K-1}{\eta-1}}$ as we saw in ~\eqref{eq:WnTauSize}) into
\begin{align} \label{eq:WRktauSplit}
W^c_{R_k,\tau}  = [ W^{c,f}_{R_k,\tau} \ \ W^{c,\overline{f}}_{R_k,\tau} ]
\end{align}
where $W^{c,f}_{R_k,\tau}$ corresponds to information that appears in a cache somewhere and that will be eventually `folded' (XORed) with other information, whereas $W_{R_k,\tau}^{c,\overline{f}}$ corresponds to information that is cached somewhere but that will not be folded with other information. The split yields
\begin{align} \label{eq:WRktauSplitSizes}
|W_{R_k,\tau}^{c,\overline{f}}| = \frac{f \alpha T-f(1-\frac{KM}{N\eta})}{\binom{K-1}{\eta}}
\end{align}
where in the above, $f \alpha T$ represents the load for each user without causing interference during the delivery phase, where $f(1-\frac{KM}{N\eta})$ is the amount of uncached information, and where $|W_{R_k,\tau}^{c,f}|=|W^c_{R_k,\tau}|-|W_{R_k,\tau}^{c,\overline{f}}|$.

We proceed to fold cached content, by creating linear combinations (XORs) from $\{W_{R_k,\tau}^{c,f}\}_{\tau\in\Psi_{\eta} \backslash \Psi_{\eta}^{(k)}}, \forall k$. We will use $P_{k,k'}(\tau)$ to be the function that replaces inside $\tau$, the entry $k'\in \tau$, with the entry $k$. As in~\cite{MN14}, the idea is that if we deliver
\begin{align} \label{eq:WRktau1}
W_{R_k,\tau}^{c,f} \oplus (\oplus_{k'\in \tau}\underbrace{W^{c,f}_{R_{k'},P_{k,k'}(\tau)}}_{\in Z_k})
\end{align}
the fact that $W^{c,f}_{R_{k'}, P_{k,k'}(\tau)} \in Z_k$, guarantees that receiver $k$ can recover $W_{R_k,\tau}^{c,f}$, while at the same time guarantees that each other user $k'\in \tau$ can recover its own desired subfile $W^{c,f}_{R_k',P_{k,k'}(\tau)} \notin Z_{k'}, \forall k' \in \tau$.

Hence delivery of each $W_{R_k,\tau}^{c,f} \oplus (\oplus_{k'\in \tau}W^{c,f}_{R_{k'},P_{k,k'}(\tau)})$ of size $|W_{R_k,\tau}^{c,f} \oplus (\oplus_{k'\in \tau}W^{c,f}_{R_{k'},P_{k,k'}(\tau)})| = |W_{R_k,\tau}^{c,f}|$ (cf.~\eqref{eq:WnTauSize}), automatically guarantees delivery of $W^{c,f}_{R_{k'},P_{k,k'}(\tau)}$ to each user $k'\in \tau$, i.e., simultaneously delivers a total of $\eta+1$ distinct subfiles (each again of size $|W^{c,f}_{R_{k'},P_{k,k'}(\tau)}| =   |W_{R_k,\tau}^{c,f}|$ bits) to $\eta+1$ distinct users. Hence \emph{any}
\begin{align} \label{eq:XpsiDef} X_{\psi} := \oplus_{k \in \psi} W^{c,f}_{R_k,\psi \backslash \{k\}}, \psi \in  \Psi_{\eta+1}\end{align}
 --- which is of the same form as in~\eqref{eq:WRktau1}, and which is referred to here as an \emph{order-($\eta+1$) folded message} --- can similarly deliver to user $k\in \psi$, her requested file $W^{c,f}_{R_k,\psi \backslash k}$, which in turn means that each order-($\eta+1$) folded message $X_{\psi}$ can deliver --- with the assistance of the side information in the caches --- a distinct, individually requested subfile, to each of the $\eta + 1$ users $k\in \psi$ ($\psi\in  \Psi_{\eta+1}$).

Thus to satisfy all requests $\{W_{R_k} \backslash Z_k \}_{k=1}^K$, the transmitter must deliver
\bit
\item uncached messages $W_{R_k}^{\overline{c}}, \ k=1,\dots,K$
\item cached but unfolded messages $\{W^{c,\overline{f}}_{R_k,\psi \backslash \{k\}}\}_{\psi \in  \Psi_{\eta+1}}, \ k=1,\dots,K$
\item and the entire set
\begin{align}\label{eq:foldedMessages}
 \mathcal{X}_\Psi := \{ X_{\psi} = \oplus_{k \in \psi} W^{c,f}_{R_k,\psi \backslash \{k\}}\}_{\psi \in  \Psi_{\eta+1}}\end{align}
consisting of
\begin{align} \label{eq:cardinalityMathcalXPsi}
|\mathcal{X}_\Psi|=\binom{K}{\eta+1}
\end{align}
folded messages of order-$(\eta+1)$, each of size (cf.~\eqref{eq:WRktauSplitSizes},\eqref{eq:WnTauSize})
\begin{align} \label{eq:XpsiSize}
|X_{\psi}| & = |W^{c,f}_{R_k,\tau}| = |W^c_{R_k,\tau}| - |W_{R_k,\tau}^{c,\overline{f}}| \nonumber\\
& = \frac{f(1-\gamma-{\alpha T)}}{\binom{K-1}{\eta}} \ \text{(bits)}.
\end{align}
\eit

\subsection{Transmission}

The transmission scheme is taken from \cite{KGZE:16}, and each transmission takes the form
\begin{align} \label{txformperfect}
\xv_{t} = \textbf{G}_{c,t} \xv_{c,t}+ \sum_{k\in \bar{\psi}}\gv_{k,t} a_{k,t}^{*}  +\sum_{k=1}^{K} \gv_{k,t} a_{k,t}
\end{align}
where  $t\in[0, T]$, where $\xv_{c,t}$ is a $K$-length vector for MAT-type symbols, where $a_{k,t}^{*}$ is the additional auxiliary symbols that carry residual interference (here we `load' this round with additional requests from the users, for the very first round, $a_{k,t}^{*} =0$) (in the above, $\bar{\psi}$ is a set of `undesired' users). In the above, each unit-norm precoder $\gv_{k,t}$ for user $k=1,2,\dots,K$, is simultaneously orthogonal to the CSI of all other channels, i.e.,
\begin{align}
\hat{\hv}_{k',t}^{T} \gv_{k,t} = 0, \ \ \forall k' \in [K] \backslash k.
\end{align}
Each precoder $\textbf{G}_{c,t}$ is defined as $\textbf{G}_{c,t} = [\gv_{c,t}, \textbf{U}_{c,t}]$, where $\gv_{c,t}$ is simultaneously orthogonal to the channel estimates of the undesired receivers, and $\textbf{U}_{c,t} \in \C^{K\times(K-1)}$ is a randomly chosen, isotropically distributed unitary matrix\footnote{Whenever possible, we will henceforth avoid going into the details of the Q-MAT scheme. Some aspects of this scheme are similar to MAT, and a main new element is that Q-MAT applies digital transmission of interference, and a double-quantization method that collects and distributes residual interference across different rounds (this is here carried by $a_{k,t}^{*}$), in a manner that allows for ZF and MAT to coexist at maximal rates. Some of the details of this scheme are `hidden' behind the choice of $\textbf{G}_{c,t}$ and behind the loading of the MAT-type symbols $\xv_{c,t}$ and additional auxiliary symbols $a_{k,t}^{*}$. The important element for the decoding part later on, will be how to load the symbols, the rate of each symbol, and the corresponding allocated power. An additional element that is hidden from the presentation here is that, while the Q-MAT scheme has many rounds, and while decoding spans more than one round, we will --- in a slight abuse of notation --- focus on describing just one round, which we believe is sufficient for the purposes of this paper here.}.

Throughout communication
\bit
\item we will allocate power such that
\begin{align}
\E\{|\xv_{c,t}|_1^2\} & \doteq  \E\{|a_{k,t}^{*}|^2\} \doteq  P , \\  \E\{|\xv_{c,t}|_{i\neq 1}^2\} & \doteq  P^{1-\alpha}, \  \E\{|a_{k,t}|^2\} \doteq P^{\alpha} \notag
\end{align}
where $|\xv_{c,t}|_i, i=1,2,\cdots,K,$ denotes scalar $i$ in vector $\xv_{c,t}$, and we will allocate rate such that
\item each $\xv_{c,t}$ carries $f(1-\alpha)$ bits per unit time,
\item each $a_{k,t}^{*}$ carries $f (1-\alpha)$ bits per unit time.
\item and each $a_{k,t}$ carries $f \alpha  $ bits per unit time.
\eit

\begin{remark}
Recall that instead of employing matrix notation, after normalization, we use the concept of signal duration $\text{dur}(\xv)$ required for the transmission of some vector $\xv $. We also note that due to time normalization, the time index $t\in [0, T]$, need not be an integer.
\end{remark}
\vspace{3pt}

For any $\alpha$, our scheme will be defined by an integer $\eta\in [\Gamma,K-1]\cap \mathbb{Z}$, which will be chosen as
\begin{align}\label{eq:etaAlpha}
\eta = \arg\max_{\eta{'}\in [\Gamma,K-1]\cap \mathbb{Z}} \{\eta{'} \ : \ \alpha_{b,\eta'}\leq \alpha\}
\end{align}
for
\begin{align}\label{eq:alphaBreak2}
{\alpha}_{b,\eta} = \frac{\eta-\Gamma}{\Gamma(H_K-H_\eta-1)+\eta} .
\end{align}
$\eta$ will define the amount of cached information that will be folded ($\{W^{c,f}_{R_k,\tau}\}_{\tau\in \Psi_{\eta} \backslash \Psi_{\eta}^{(k)}}$ ), and thus also the amount of cached information that will not be folded ($\{W^{c,\overline{f}}_{R_k,\tau}\}_{\tau\in \Psi_{\eta}\backslash \Psi_{\eta}^{(k)}}$ ) and which will be exclusively carried by the different $a_{k,t}$. In all cases,
\bit
\item all of $\{X_{\psi}\}_{\psi \in \Psi_{\eta+1}}$ (which are functions of the cached-and-to-be-folded $\{W^{c,f}_{R_k,\tau}\}_{\tau\in \Psi_{\eta} \backslash \Psi_{\eta}^{(k)}}$) will be exclusively carried by $\xv_{c,t}, \ t\in[0, T]$, while
\item all of the uncached $W^{\overline{c}}_{R_k}$ (for each $k=1,\dots,K$) and all of the cached but unfolded $\{W^{c,\overline{f}}_{R_k,\tau}\}_{\tau\in\Psi_{\eta} \backslash \Psi_{\eta}^{(k)}}$ will be exclusively carried by $a_{k,t}, t\in[0,T]$.
\eit

\emph{Transmission of $\{X_{\psi}\}_{\psi \in  \Psi_{\eta+1}}$:} From \cite{KGZE:16}, we know that the transmission relating to $\xv_{c,t}$ can be treated independently from that of $a_{k,t}$, simply because --- as we will further clarify later on --- the $a_{k,t}$ do not actually interfere with decoding of $\xv_{c,t}$, as a result of the scheme, and as a result of the chosen power and rate allocations which jointly adapt to the CSIT quality $\alpha$. For this reason, we can treat the transmission of $\xv_{c,t}$ separately.

Hence we first focus on the transmission of $\{X_{\psi}\}_{\psi \in  \Psi_{\eta+1}}$, which will be sent using $\xv_{c,t}, \ t\in[0,T]$ using the last $K-\eta$ phases of the QMAT algorithm in~\cite{KGZE:16} corresponding to having the ZF symbols $a_{k,t}$ set to zero. For ease of notation, we will label these phases starting from phase $\eta+1$ and terminating in phase $K$. The total duration is the desired $T$. Each phase $j = \eta+1, \dots,K$ aims to deliver order-$j$ folded messages (cf.~\eqref{eq:foldedMessages}), and will do so gradually: phase $j$ will try to deliver (in addition to other information) $N_j := (K-j+1)\binom{K}{j}$ order-$j$ messages which carry information that has been requested by $j$ users, and in doing so, it will generate $N_{j+1} := j\binom{K}{j+1}$ signals that are linear combinations of received signals from $j+1$ different users, and where these $N_{j+1}$ signals will be conveyed in the next phase $j+1$. During the last phase $j=K$, the transmitter will send fully common symbols that are useful and decoded by all users, thus allowing each user to go back and retroactively decode the information of phase $j=K-1$, which will then be used to decode the information in phase $j=K-2$ and so on, until they reach phase $j = \eta+1$ (first transmission phase) which will complete the task. We proceed to describe these phases. We will use $T_j$ to denote the duration of phase $j$.

\emph{Phase $\eta+1$:} In this first phase of duration $T_{\eta+1}$, the information in $\{X_{\psi}\}_{\psi \in \Psi_{\eta+1}}$ is delivered by $\xv_{c,t}, \ t\in[0,T_{\eta+1}] $, which can also be rewritten in the form of a sequential transmission of shorter-duration $K$-length vectors
\begin{align}\label{eq:vectorfirstphase}
\xv_{\psi} = [x_{\psi,1}, \dots, x_{\psi,K-\eta}, 0, \dots, 0]^{T}
\end{align}
for different $\psi$, where each vector $\xv_{\psi}$ carries exclusively the information from each $X_{\psi}$, and where this information is uniformly split among the $K-\eta$ independent scalar entries $x_{\psi,i}, \ i=1,\dots,K-\eta$, each carrying
\begin{align} \label{eq:SizeScalar_x_psi}
\frac{|X_{\psi}|}{(K-\eta)} = \frac{f(1-\gamma-\alpha T)}{\binom{K-1}{\eta}(K-\eta)}
\end{align}
bits (cf.~\eqref{eq:XpsiSize}).
Hence, given that the allocated rate for $\xv_{c,t}$ (and thus the allocated rate for each $\xv_{\psi}$) is $(1-\alpha)f$, we have that the duration of each $\xv_{\psi}$ is
\begin{align} \label{eq:durationVector_XPsi}
\text{dur}(\xv_{\psi}) = \frac{|X_{\psi}|}{(K-\eta)(1-\alpha)f}.
\end{align}
Given that $|\mathcal{X}_\Psi|=\binom{K}{\eta+1}$, then
\begin{align} \label{eq:durationcal1}
T_{\eta+1} = \binom{K}{\eta+1} \text{dur}(\xv_{\psi}) =\frac{ \binom{K}{\eta+1}  |X_{\psi}|}{(K-\eta)(1-\alpha)f}.
\end{align}

After each transmission of $\xv_{\psi}$, each user $k\in[K]$ receives (in addition to information originating from $a_{k,t}$ which will be treated as noise and thus neglected for now), a linear combination $L_{\psi,k}$ of the transmitted $K-\eta$ symbols $x_{\psi,1}, x_{\psi,2}, \dots, x_{\psi,K-\eta}$.
Next the transmitter will send an additional $K-\eta-1$ signals $L_{\psi,k'}, \ k'\in[K]\backslash \psi$ (linear combinations of $x_{\psi,1}, x_{\psi,2}, \dots, x_{\psi,K-\eta}$ as received --- up to noise level --- at each user $k'\in [K] \backslash \psi$) which will help each user $k\in \psi$ resolve the already sent $x_{\psi,1}, x_{\psi,2}, \dots, x_{\psi,K-\eta}$. This will be done in the next phase $j=\eta+2$.

\emph{Phase $\eta+2$:} The challenge now is for signals $\xv_{c,t}, \ t \in (T_{\eta+1},T_{\eta+1}+T_{\eta+2}] $ to convey all the messages of the form
\[L_{\psi,k'}, \ \forall k'\in[K]\backslash \psi, \ \forall \psi\in \Psi_{\eta+1} \]
to each receiver $k\in \psi$. Note that each of the above linear combinations, is now --- during this phase --- available (up to noise level) at the transmitter. Let
\begin{align}
\Psi_{\eta+2} = \{\psi\in [K] \ : \  |\psi|=\eta+2  \}
\end{align}
and consider for each $\psi\in \Psi_{\eta+2}$, a transmitted vector
\[\xv_\psi = [x_{\psi,1}, \dots, x_{\psi,K-\eta-1}, 0, \dots, 0]^{T}\]
which carries the contents of $\eta+1$ different linear combinations $f_i(\{L_{\psi\backslash\{k\},k}\}_{k\in \psi}), i=1,\dots,\eta+1$ of the $\eta+2$ elements $\{L_{\psi\backslash\{k\},k}\}_{\forall k\in \psi}$ created by the transmitter. The linear combination coefficients defining each linear-combination function $f_i$, are predetermined and known at each receiver. The transmission of $\{\xv_{\psi}\}_{\forall \psi \in \Psi_{\eta+2}}$ is sequential.

It is easy to see that there is a total of $(\eta+1)\binom{K}{\eta+2}$ symbols of the form $f_i (\{L_{\psi\backslash\{k\},k}\}_{k\in \psi}), i=1,\dots,\eta+1, \ \psi\in \Psi_{\eta+2}$, each of which can be considered as an order-$(\eta+2)$ signal intended for $\eta+2$ receivers in $\psi$. Using this, and following the same steps used in phase $\eta+1$, we calculate that
\begin{align} \label{eq:durationPhase2}
T_{\eta+2} = \binom{K}{\eta+2}\text{dur}(\xv_{\psi}) = T_{\eta+1} \frac{\eta+1}{\eta+2}.
\end{align}

We now see that for each $\psi$, each receiver $k \in \psi$ recalls their own observation $L_{\psi \backslash \{k\}, k}$ from the previous phase, and removes it from all the linear combinations $\{ f_i (\{L_{\psi\backslash\{k\},k}\}_{\forall  k\in \psi})\}_{i=1,\dots,\eta+1}$, thus now being able to acquire the $\eta+1$ independent linear combinations $\{L_{\psi \backslash \{k'\}, k'}\}_{\forall  k' \in \psi \backslash \{k\}}$. The same holds for each other user $k'\in \psi$.

After this phase, we use $L_{\psi,k}, \psi \in \Psi_{\eta+2}$ to denote the received signal at receiver $k$. Like before, each receiver $k, k\in \psi$ needs $K-\eta-2$ extra observations of $x_{\psi,1}, \dots, x_{\psi,K-\eta-1}$ which will be seen from $L_{\psi,k'},\forall k'\notin \psi$, which will come from order-$(\eta+3)$ messages that are created by the transmitter and which will be sent in the next phase.

\emph{Phase $j$ $(\eta+3 \leq j \leq K)$:}
Generalizing the described approach to any phase $j\in[\eta+3,\dots,K]$, we will use $\xv_{c,t}, \ t \in [\sum_{i=\eta+1}^{j-1}T_{i},\sum_{i=\eta+1}^{j}T_{i} ] $ to convey all the messages of the form
\[L_{\psi,k'}, \ \forall k'\in[K]\backslash \psi, \ \forall \psi\in \Psi_{j-1} \]
to each receiver $k\in \psi$. For each
\begin{align}
\psi\in \Psi_{j} := \{\psi\in [K] \ : ~ |\psi|=j  \}
\end{align}
each transmitted vector
\[\xv_\psi = [x_{\psi,1}, \dots, x_{\psi,K-j-1}, 0, \dots, 0]^{T}\]
will carry the contents of $j-1$ different linear combinations $f_i(\{L_{\psi\backslash\{k\},k}\}_{k\in \psi}), i=1,\dots,j-1$ of the $j$ elements $\{L_{\psi\backslash\{k\},k}\}_{\forall k\in \psi}$ created by the transmitter. After the sequential transmission of $\{\xv_{\psi}\}_{\forall \psi \in \Psi_{j}}$, each receiver $k$ can obtain the $j-1$ independent linear combinations $\{L_{\psi \backslash \{k'\}, k'}\}_{\forall  k' \in \psi \backslash \{k\}}$. The same holds for each other user $k'\in \psi$.
As with the previous phases, we can see that
\begin{align} \label{eq:durationPhaseJ}
T_j = T_{\eta+1} \frac{\eta+1}{j}, \ j= \eta+3,\dots,K.
\end{align}
This process terminates with phase $j = K$, during which each \[\xv_\psi = [x_{\psi,1}, 0 , 0, \dots, 0]^{T}\] carries a single scalar that is decoded easily by all. Based on this, backwards decoding will allow for users to retrieve $\{ X_{\psi}\}_{\psi \in \Psi_{\eta+1}}$. This is described immediately afterwards. In treating the decoding part, we briefly recall that each $a_{k,t}, \ k=1,\dots,K$ carries (during period $t\in[0,T]$), all of the uncached $W^{\overline{c}}_{R_k}$ and all of the unfolded $\{W^{c,\overline{f}}_{R_k,\tau}\}_{\tau\in\Psi_{\eta} \backslash \Psi_{\eta}^{(k)}}$.

\subsection{Decoding}

The whole transmission lasts $K-\eta$ phases. For each phase $j, j=\eta+1,\cdots,K$ and the corresponding $\psi$, the received signal $y_{k,t}, \ t \in [\sum_{i=\eta+1}^{j-1}T_{i},\sum_{i=\eta+1}^{j}T_{i} ]$ of desired user $k~(k \in \psi)$  takes the form
\begin{align}
y_{k,t} = \underbrace{\overbrace{\hv_{k,t}^{T} \textbf{G}_{c,t} \xv_{c,t}}^{\text{rate} \ 1-\alpha} }_{L_{\psi,k},  \ \text{power} \ \doteq \ P}+ \underbrace{\overbrace{\hv_{k,t}^{T} \sum^K_{k \in \bar{\psi}}\gv_{k,t}  a_{k,t}^{*}}^{\text{rate} \ 1-\alpha} }_{\doteq \ P^{1-\alpha}}+  \underbrace{\overbrace{\hv_{k,t}^{T} \gv_{k,t} a_{k,t}}^{\text{rate} \ \alpha }}_{P^{\alpha}}
\end{align}
The received signal $y_{k',t}$ of undesired user $k'~(k' \in [K]\backslash \psi)$ takes the form
\begin{align}
y_{k',t} = \underbrace{\overbrace{\hv_{k',t}^{T} \sum^K_{k' \in \bar{\psi}}\gv_{k',t}  a_{k',t}^{*}}^{\text{rate} \ 1-\alpha} }_{\ \text{power} \ \doteq \ P} + \underbrace{\overbrace{\hv_{k',t}^{T} \textbf{G}_{c,t} \xv_{c,t}}^{\text{rate} \ 1-\alpha} }_{L_{\psi,k'},  \ \ \doteq \ P^{1-\alpha}}+ \underbrace{\overbrace{\hv_{k',t}^{T} \gv_{k',t} a_{k',t}}^{\text{rate} \ \alpha }}_{P^{\alpha}}
\end{align}
where in both cases, we ignore the Gaussian noise and the ZF-related noise up to $P^{0}$. In addition to somehow send $L_{\psi,k'}$ to the next MAT phase, as we see in~\cite{KGZE:16}, after each phase, $L_{\psi,k'}$ is first quantized with $(1-2\alpha) \log P$ bits, which results in a residual quantizaton noise $n_{\psi,k'}$ with power scaling as $P^{\alpha}$. Then, the transmitter quantizes the quantization noise $n_{\psi,k'}$ with an additional $\alpha \log P$ bits, which will be carried by the auxiliary data symbols $a_{k',t}^{*}$ in the corresponding phase in the next round (note: additional requests from the users). In this way, we can see that the `common' signal $\xv_{c,t}$ can be decoded with the assistance of an auxiliary data symbol from the next round. After this, each user $k$ will remove $\hv_{k,t}^{T} \textbf{G}_{c,t} \xv_{c,t}$ from their received signals, and readily decode their private symbols $a_{k,t}, \ t\in[0,T]$, thus allowing for retrieval of their own unfolded $\{W^{c,\overline{f}}_{R_k,\psi \backslash \{k\}}\}_{\psi \in \Psi_{\eta+1}}$  and uncached $W_{R_k}^{\overline{c}}$.
In terms of decoding the common information, as discussed above, each receiver $k$ will perform a backwards reconstruction of the sets of overheard equations
\[\ba{c}
\{L_{\psi,k'}, \ \forall k'\in[K]\backslash \psi\}_{\forall \psi\in \Psi_{K}} \\ \downarrow \\ \{L_{\psi,k'}, \ \forall k'\in[K]\backslash \psi\}_{\forall \psi\in \Psi_{K-1}} \\ \vdots \\ \downarrow \\ \{L_{\psi,k'}, \ \forall k'\in[K]\backslash \psi\}_{\forall \psi\in \Psi_{\eta+2}}  \ea
\]
until phase $\eta+2$. At this point, each user $k$ has enough observations to recover the original $K-\eta$ symbols $x_{\psi,1}, x_{\psi,2}, \dots, x_{\psi,K-\eta}$ that fully convey $X_{\psi}$, hence each user $k$ can reconstruct their own set $\{W^{c,f}_{R_k,\psi \backslash \{k\}}\}_{\psi \in  \Psi_{\eta+1}}$ which, combined with the information from the $a_{k,t}, \ t=[0,T]$ allow for each user $k$ to reconstruct $\{W^c_{R_k,\psi \backslash \{k\}}\}_{\psi \in \Psi_{\eta+1}}$ which is then combined with $Z_{k}$ to allow for reconstruction of the requested file $W_{R_k}$.

\subsection{Calculation of $T$}
To calculate $T$, we recall from \eqref{eq:durationPhaseJ} that
\begin{align}
T &= \sum \limits_{j=\eta+1}^{K} T_j = T_{\eta+1}\sum \limits_{j=\eta+1}^{K} \frac{\eta+1}{j} \nonumber \\ &= (\eta+1)(H_K-H_\eta)T_{\eta+1} \label{eq:durationcal2}
\end{align}
which combines with~\eqref{eq:SizeScalar_x_psi} and~\eqref{eq:durationcal1} to give
\begin{align} \label{eq:proofT1}
T = \frac{(K-\Gamma)(H_K-H_\eta)}{(K-\eta)+\alpha(\eta+K(H_K-H_\eta-1))}
\end{align}
as stated in Theorem~\ref{thm:bigGammaBest}. The bound by $T=1-\gamma$ seen in the theorem, corresponds to the fact that the above expression~\eqref{eq:proofT1} applies, as is, only when $\alpha\leq \alpha_{b,K-1}  = \frac{K(1-\gamma)-1}{(K-1)(1-\gamma)}$ which corresponds to $\eta = K-1$ (where $X_\psi$ are fully common messages, directly desired by all), for which we already get the best possible $T=1-\gamma$, and hence there is no need to go beyond $\alpha=\frac{K(1-\gamma)-1}{(K-1)(1-\gamma)}$.

\section{Conclusions \label{sec:conclusions}}

This work studied the previously unexplored interplay between coded-caching and CSIT feedback quality and timeliness. This is motivated by the fact that CSIT and coded caching are two powerful ingredients that are hard to obtain, and by the fact that these ingredients are intertwined in a synergistic and competing manner.
In addition to the substantial cache-aided DoF gains revealed here, the results suggest the interesting practical ramification that distributing predicted content `during the night', can offer continuous amelioration of the load of predicting and disseminating CSIT during the day.

\section{Appendix\label{sec:additionalProofs}}
\subsection{Lower bound on $T^*$ (proof of Lemma~\ref{lem:outer})\label{sec:lower}}

This part draws from the bound in~\cite{MN14}, and it is similar to that in \cite{ZEsynergy:16} which deals with the case of $\alpha = 0$.
To lower bound $T$, we consider the easier problem where we want to serve $s\leq K $ different files to $s$ users, each with access to all caches.
We also consider that we repeat this (easier) last experiment $\lfloor \frac{N}{s} \rfloor$ times, thus spanning a total duration of $T \lfloor \frac{N}{s} \rfloor$  (and up to $\lfloor \frac{N}{s} \rfloor s$ files delivered). At this point, we transfer to the equivalent setting of the $s$-user MISO BC with delayed CSIT and imperfect current CSIT, and a side-information multicasting link to the receivers, of capacity $d_m$ (files per time slot). Under the assumption that in this latter setting, decoding happens at the end of communication, and once we set
\begin{align} \label{eq:fromdmToCaching} d_m T \lfloor \frac{N}{s} \rfloor = sM \end{align}
(which guarantees that the side information from the side link, throughout the communication process, matches the maximum amount of information in the caches), we have that \begin{align}
T \lfloor \frac{N}{s}  \rfloor d^{'}_{\Sigma}(d_m) \geq \lfloor \frac{N}{s} \rfloor s
\end{align}
where $d^{'}_{\Sigma}(d_m)$ is any sum-DoF upper bound on the above $s$-user MISO BC channel with delayed CSIT and the aforementioned side link.
Using the bound
\[d^{'}_{\Sigma}(d_m) = s \alpha  + \frac{s}{H_s} (1-\alpha+d_m)  \]
from Lemma~\ref{lem:lowerSecond}, and applying \eqref{eq:fromdmToCaching}, we get
\begin{align}
T \lfloor \frac{N}{s}  \rfloor \bigl( s\alpha + \frac{s}{H_s}(1-\alpha + \frac{sM}{T \lfloor \frac{N}{s}  \rfloor }) \bigr)\geq \lfloor \frac{N}{s} \rfloor s
\end{align}
and thus that
\begin{align}
T \geq \frac{1}{(H_s \alpha+1-\alpha)} (H_s -\frac{Ms}{\lfloor \frac{N}{s}\rfloor })
\end{align} 
which implies a lower bound on the original $s$-user problem.
Maximization over all $s$, gives the desired bound on the optimal $T^*$
\begin{align}
T^* \geq \mathop {\text{max}}\limits_{s\in \{1, \dots, \lfloor \frac{N}{M} \rfloor \}}  \frac{1}{(H_s \alpha+1-\alpha)} (H_s -\frac{Ms}{\lfloor \frac{N}{s}\rfloor})
\end{align}
required for the original $K$-user problem. This concludes the proof.

\subsection{Bounding the sum-DoF of the $s$-user MISO BC, with delayed CSIT, $\alpha$-quality current CSIT, and additional side information\label{sec:lowerSecond}}
We begin with the statement of the lemma, which we prove immediately below.
\begin{lemma} \label{lem:lowerSecond}
For the $s$-user MISO BC, with delayed CSIT, $\alpha$-quality current CSIT, and an additional parallel side-link of capacity that scales as $d_m \log P$, the sum-DoF is upper bounded as
\begin{align}
\label{eq:lowerSecond}
d_{\Sigma}(d_m) \leq  s \alpha  + \frac{s}{H_s} (1-\alpha+d_m).
\end{align}
\end{lemma}

\vspace{3pt}
\begin{proof}
Our proof traces the proof of \cite{CYOG:14}, adapting for the additional $\alpha$-quality current CSIT.

Consider a permutation $\pi$ of the set $\mathcal{E}= \{1,2,\cdots,s\}$. For any user $k , k \in \mathcal{E}$, we provide the received signals $y_k^{[n]}$ as well as the message $W_k$ of user $k$ to user $k+1, k+2, \cdots, s$. We define the following notation
\[\Omega^{[n]} := \{\hv_k^{[n]}\}_{k=1}^{s},~~ \hat{\Omega}^{[n]} := \{\hat\hv_k^{[n]}\}_{k=1}^{s}, ~~\mathcal{U}^{[n]} := \{\Omega^{[n]}, \hat{\Omega}^{[n]}\},\]
\[\hv_k^{[t]} := \{\hv_k^{(i)}\}_{i=1}^{t} ,~~ y_k^{[t]}:= \{y_k^{(i)}\}_{i=1}^{t}, t=1,2,\cdots,n,\]
\[ W_{[k]} := \{W_1,W_2,\cdots,W_k\},~~y_{[k]}^{[n]} := \{y_1^{[n]},y_2^{[n]},\cdots,y_k^{[n]}\}.\]
Then for $k =1,2,\cdots,s$, we have
\begin{align}
& n (R_k-\epsilon_n) \notag \\
                   &\leq I(W_k; y_{[k]}^{[n]},y_0^{[n]}, W_{[k-1]} | \mathcal{U}^{[n]}) \label{eq:DoFbound1} \\
                   & = I(W_k; y_{[k]}^{[n]},y_0^{[n]}| W_{[k-1]}, \mathcal{U}^{[n]})  \label{eq:DoFbound2} \\
                   & = I(W_k; y_{[k]}^{[n]}| W_{[k-1]}, \mathcal{U}^{[n]}) +   I(W_k; y_0^{[n]}| y_{[k]}^{[n]}, W_{[k-1]}, \mathcal{U}^{[n]}) \notag \\
								   & = h(y_{[k]}^{[n]}|W_{[k-1]}, \mathcal{U}^{[n]}) - h(y_{[k]}^{[n]}| W_{[k]}, \mathcal{U}^{[n]}) \notag \\
									 & ~~ + h(y_0^{[n]}| y_{[k]}^{[n]}, W_{[k-1]}, \mathcal{U}^{[n]}) - h(y_0^{[n]}|y_{[k]}^{[n]}, W_{[k]}, \mathcal{U}^{[n]}) \label{eq:DoFbound7}
\end{align}
where \eqref{eq:DoFbound1} follows from Fano's inequality, where \eqref{eq:DoFbound2} holds due to the fact that the messages are independent, and where the last two steps use the basic chain rule. Note that $W_0 = 0$.

\begin{align}
&\sum_{k=1}^{s-1} \big( \frac{h(y_{[k+1]}^{[n]}|W_{[k]}, \mathcal{U}^{[n]})}{k+1}  - \frac{h(y_{[k]}^{[n]}| W_{[k]}, \mathcal{U}^{[n]})}{k} \big) \notag \\
                 &=\sum_{t=1}^{[n]} \sum_{k=1}^{s-1} \big( \frac{h( y_{1}^{(t)},\cdots,y_{k+1}^{(t)}|y_{1}^{[t-1]},\cdots,y_{k+1}^{[t-1]}, W_{[k]}, \mathcal{U}^{[n]})}{k+1}  \notag \\
								 &~~- \frac{h(y_{1}^{(t)},\cdots,y_{k}^{(t)}|y_{1}^{[t-1]},\cdots,y_{k}^{[t-1]}, W_{[k]}, \mathcal{U}^{[n]})}{k} \big) \label{eq:DoFbound3}   \\
		             &=\sum_{t=1}^{[n]} \sum_{k=1}^{s-1} \big( \frac{h( y_{1}^{(t)},\cdots,y_{k+1}^{(t)}|y_{1}^{[t-1]},\cdots,y_{k+1}^{[t-1]}, W_{[k]}, \mathcal{U}^{[t]})}{k+1}  \notag \\
								 &~~- \frac{h(y_{1}^{(t)},\cdots,y_{k}^{(t)}|y_{1}^{[t-1]},\cdots,y_{k}^{[t-1]}, W_{[k]}, \mathcal{U}^{[t]})}{k} \big) \label{eq:DoFbound4} \\
		             &\leq \sum_{t=1}^{[n]} \sum_{k=1}^{s-1} \big( \frac{h( y_{1}^{(t)},\cdots,y_{k+1}^{(t)}|y_{1}^{[t-1]},\cdots,y_{k+1}^{[t-1]}, W_{[k]}, \mathcal{U}^{[t]})}{k+1}   \notag \\
								 &~~- \frac{h(y_{1}^{(t)},\cdots,y_{k}^{(t)}|y_{1}^{[t-1]},\cdots,y_{k+1}^{[t-1]}, W_{[k]}, \mathcal{U}^{[t]})}{k} \big) \label{eq:DoFbound5}	\\
		            &\leq \sum_{t=1}^{[n]} \sum_{k=1}^{s-1}  \frac{1}{k+1} \alpha \log P + 	n \cdot o(\log P)	\label{eq:DoFbound6} 	\\
						   & \leq  n (H_s-1) \alpha \log P  + 	n \cdot o(\log P)	
\end{align}
where \eqref{eq:DoFbound3} follows from the linearity of the summation, where \eqref{eq:DoFbound4} holds since the received signal is independent of the future channel state information, where \eqref{eq:DoFbound5} uses the fact that conditioning reduces entropy, and where \eqref{eq:DoFbound6} is from the fact that
Gaussian distribution maximizes differential entropy under the covariance constraint and from \cite[Lemma~2]{KYG:13}.
From \eqref{eq:DoFbound7}, we then have
\begin{align}
&\sum_{k=1}^{s} \frac{n (R_k-\epsilon_n) }{k} \notag \\
								 	&\leq \sum_{k=1}^{s-1} \big( \frac{h(y_{[k+1]}^{[n]}|W_{[k]}, \mathcal{U}^{[n]})}{k+1}  - \frac{h(y_{[k]}^{[n]}| W_{[k]}, \mathcal{U}^{[n]})}{k} \big) \notag \\
									  &~~+ h(y_1^{[n]}|\mathcal{U}^{[n]})-\frac{1}{s} h(y_{[s]}^{[n]}|W_{[s]}, \mathcal{U}^{[n]}) \notag \\
										&~~+ \!\! \sum_{k=1}^{s-1} \big( \frac{H(y_0^{[n]}|y_{[k+1]}^{[n]}, W_{[k]}, \mathcal{U}^{[n]})}{k+1} \! - \! \frac{H(y_0^{[n]}|y_{[k]}^{[n]}, W_{[k]}, \mathcal{U}^{[n]})}{k}  \big) \notag \\
										&~~+ H(y_0^{[n]}|y_1^{[n]}, \mathcal{U}^{[n]})-\frac{1}{s} H (y_0^{[n]}|y_{[s]}^{[n]},W_{[s]}, \mathcal{U}^{[n]})	\notag \\
										& \leq n (H_s-1) \alpha \log P   + \underbrace{h(y_1^{[n]}|\mathcal{U}^{[n]})}_{\leq n \log P} + \underbrace{H(y_0^{[n]}|y_1^{[n]}, \mathcal{U}^{[n]})}_{\leq n \cdot d_m \log P} \notag \\
										&~~+ \sum_{k=1}^{s-1} \big( (\frac{1}{k+1}-\frac{1}{k}) H(y_0^{[n]}|y_{[k]}^{[n]}, W_{[k]}, \mathcal{U}^{[n]} \big) + n \cdot o(\log P)	  \notag \\
										&\leq n (H_s-1) \alpha \log P   +  n  \log P +n \cdot  d_m \log P + n \cdot o(\log P).
\end{align}
Dividing by $n  \log P$ and letting  $P \rightarrow \infty$ gives
\begin{align}
 \sum_{k=1}^{s} \frac{d_k}{k} \leq  (H_s-1) \alpha    + 1 + d_m
\end{align}
which implies that
\begin{align}
d_{\Sigma}(d_m) \leq  s \alpha  + \frac{s}{H_s} (1-\alpha+d_m)
\end{align}
which completes the proof of~Lemma~\ref{lem:lowerSecond}.

\end{proof}

\subsection{Bounding the gap between the achievable $T$ and the optimal $T^*$ \label{sec:gapCalculation}}
Our aim here is to show that \[  \frac{T(\gamma,\alpha > 0)}{T^*(\gamma,\alpha > 0)}<4 \]
and we will do so by showing that the above gap is smaller than the gap we calculated in \cite{ZEsynergy:16} for $\alpha = 0$, which was again bounded above by 4.
For this, we will use the expression\footnote{We note that the here derived upper bound on the gap corresponding to the $T$ in Theorem~\ref{thm:bigGamma}, automatically applies as an upper bound to the gap corresponding to the $T$ from Theorem~\ref{thm:bigGammaBest}, because the latter $T$ is smaller than the former.}
\begin{align} \label{eq:gapProofT1}
T(\gamma,\alpha > 0) = \frac{(1-\gamma)(H_K-H_{\Gamma})}{\alpha(H_K-H_{\Gamma})+(1-\alpha)(1-\gamma)}
\end{align}
from Theorem~\ref{thm:bigGamma}, and the expression
\begin{align}
T^*(\gamma,\alpha > 0) \geq \mathop {\text{max}}\limits_{s\in \{1, \dots, \lfloor \frac{N}{M} \rfloor \}} \frac{1}{(H_s \alpha+1-\alpha)} (H_s -\frac{Ms}{\lfloor \frac{N}{s} \rfloor})
\end{align}
from Lemma~\ref{lem:outer}. Hence we have
\begin{align}
\frac{T}{T^*} & \leq \frac{\frac{(1-\gamma)(H_K-H_{K \gamma})}{\alpha(H_K-H_{K \gamma})+(1-\alpha)(1-\gamma)}}{\max\limits_{s\in \{1, \dots, \lfloor \frac{N}{M} \rfloor\}} \frac{1}{(H_s \alpha+1-\alpha)} (H_s -\frac{Ms}{\lfloor \frac{N}{s} \rfloor})} \label{eq:bound20} \\
& \leq \underbrace{\frac{\frac{(1-\gamma)(H_K-H_{K \gamma})}{\alpha(H_K-H_{K \gamma})+(1-\alpha)(1-\gamma)}}{\frac{1}{(H_{s_c} \alpha+1-\alpha)}(H_{s_c} -\frac{Ms_c}{\lfloor \frac{N}{s_c} \rfloor})}}_{g(s_c,\gamma)}\label{eq:bound20b}
\end{align}
where $s=s_c\in \{1, \dots, \lfloor \frac{N}{M} \rfloor\}$, but where this $s_c$ will be chosen here to be exactly the same as in the case of $\alpha = 0$. This will be useful because, for that case of $\alpha = 0$, we have already proved that the same specific $s_c$ guarantees that
\begin{align}
\frac{H_{K}-H_{K\gamma}}{H_{s_c} -\frac{Ms_c}{\lfloor \frac{N}{s_c} \rfloor}} < 4 \label{bound21}
\end{align}
for the appropriate ranges of $\gamma$. This will apply towards bounding~\eqref{eq:bound20b}.

The proof is broken in two cases, corresponding to $\gamma\in[\frac{1}{36} ,\frac{K-1}{K}]$, and $\gamma\in[0 ,\frac{1}{36}]$.

\subsubsection{Case 1 ($\alpha > 0,\gamma\in[\frac{1}{36} ,\frac{K-1}{K}]$)}
As when $\alpha = 0$ (cf.~\cite{ZEsynergy:16}), we again set $s=1$, which reduces~\eqref{eq:bound20b} to
\[\frac{T(\alpha>0,\gamma)}{T^*(\alpha>0,\gamma)} \leq \frac{\frac{(1-\gamma)(H_K-H_{K \gamma})}{\alpha(H_K-H_{K \gamma})+(1-\alpha)(1-\gamma)}}{1-\gamma}.\]
For this case --- when $\alpha$ was zero, and when we chose the same $s = 1$ --- we have already proved that
\[\frac{T(\alpha=0,\gamma)}{1-\gamma}<4. \]
As a result, since $T(\alpha>0,\gamma)<T(\alpha=0,\gamma)$, and since $1-\gamma\leq T^*$, we conclude that
\[\frac{T(\alpha>0,\gamma)}{T^*}<4, \ \gamma\in[\frac{1}{36} ,\frac{K-1}{K}]\]
which completes this part of the proof.

\subsubsection{Case 2 ($\alpha > 0, \gamma\in[0,\frac{1}{36}]$)}
Going back to~\eqref{eq:bound20b}, we now aim to bound
\begin{align} \label{eq:def_g1}
g(s_c,\gamma):=\frac{\frac{(1-\gamma)(H_K-H_{K \gamma})}{\alpha(H_K-H_{K \gamma})+(1-\alpha)(1-\gamma)}}{\frac{1}{(H_{s_c} \alpha+1-\alpha)}(H_{s_c} -\frac{Ms_c}{\lfloor \frac{N}{s_c} \rfloor})} < 4.
\end{align}
We already know from the case of $\alpha = 0$ (cf.~\eqref{bound21}) that
\begin{align}
\frac{H_K-H_{K\gamma}}{H_{s_c} -\frac{Ms_c}{\lfloor \frac{N}{s_c} \rfloor}} <4 \label{bound22-aa}
\end{align}
holds.
Hence we will prove that
\begin{align} \label{bound22-bb}
g(s_c,\gamma) \leq \frac{H_K-H_{K\gamma}}{H_{s_c} -\frac{Ms_c}{\lfloor \frac{N}{s_c} \rfloor}}
\end{align}
to guarantee the bound.
We note that~\eqref{bound22-bb} is implied by
\begin{align}
H_{s_c} \leq \frac{H_K-H_{K \gamma}}{1-\gamma} \label{bound22}
\end{align}
which is implied by
\begin{align}
\log(s_c) \leq \frac{\log(1/ \gamma)}{1-\gamma}-\epsilon_6,  \ \epsilon_6=H_6-\log(6) \label{bound23}
\end{align}
because $H_{s_c} \leq \log(s_c)+\epsilon_6, \forall s_c \geq 6,\forall \gamma \in [0, \frac{1}{36}], \forall K$.
Furthermore~\eqref{bound23} is implied by
\begin{align}
\frac{1}{2} \log(\frac{1}{\gamma}) \leq \frac{\log(1/ \gamma)}{1-\gamma}-\epsilon_6 \label{bound24}
\end{align}
because $\gamma \in [\frac{1}{(s_c+1)^2}, \frac{1}{s_c^2}]$ means that $s_c \leq \sqrt{\frac{1}{\gamma}}$. Since $\frac{1}{1-\gamma} \geq 1$, then \eqref{bound24} is implied by
\begin{align}
\frac{1}{2} \log(\frac{1}{\gamma}) \leq \log(\frac{1}{\gamma})-\epsilon_6 \label{bound25}.
\end{align}
It is obvious that~\eqref{bound25} holds since $\gamma \leq \frac{1}{36}$. Towards this, by proving~\eqref{bound25}, we guarantee \eqref{eq:def_g1} and the desired bound. This completes the proof.

\nocite{ZE:15}

\bibliographystyle{IEEEtran}
\bibliography{IEEEabrv,final_refs}

\end{document}